\documentclass[prb,showpacs,byrevtex,amsmath,amssymb,nofootinbib,preprint]{revtex4}

\usepackage{graphicx}
\usepackage{dcolumn}
\usepackage{bm}

\begin{document}
\preprint{PREPRINT}

\title[Short Title]{Macroion adsorption: The crucial role of excluded volume and coions}

\author{Ren\'e Messina}
\email{messina@thphy.uni-duesseldorf.de}
\affiliation
{Institut f\"ur Theoretische Physik II,
Heinrich-Heine-Universit\"at D\"usseldorf,
Universit\"atsstrasse 1,
D-40225 D\"usseldorf,
Germany}

\date{\today}

\begin{abstract}
The adsorption of charged colloids (macroions) onto an oppositely charged planar substrate
is investigated theoretically.
Taking properly into account the finite size of the macroions , unusual behaviors are reported.
It is found that the role of the coions 
(the little salt-ions carrying the same sign of charge as that of the substrate) is
crucial to understand the mechanisms involved in the process of macroion adsorption. 
In particular, the coions can accumulate near the substrate's surface and lead to
a counter-intuitive {\it surface charge amplification}. 
\end{abstract}
\pacs{68.43.De, 68.08.-p, 82.70.Dd, 61.20.Gy}
\maketitle

\section{Introduction}

Whereas the bulk behavior of homogeneous 
(charged \cite{Loewen_AnnRevPhyChem_2000,Levin_RepProgPhys_2002} 
and uncharged \cite{Hansen_Book_1990}) 
colloidal suspensions is rather well understood,
the situation for its inhomogeneous counterpart,
such as that emerging in an adsorption process, is less clear.
Potential applications of adsorption of charged colloidal particles 
(macroions) can vary from technological processes such as surface coating 
\cite{Decher_1997} to biological material problems. \cite{Kawaguchi_Prog_Polym_Sci_2000}
From a fundamental point of view, the tremendous long-ranged Coulomb 
interaction that sets in represents a formidable theoretical challenge.
Consequently, a deeper understanding of the phenomenon of 
macroion adsorption is justified and needed.    
 
On one hand, experiments \cite{Bu_Langmuir_2006,Luo_Science_2006}
and the well known  mean field Gouy-Chapman theory 
\cite{Gouy_JPhys_1910,Chapman_PhilMag_1913} 
seem to nicely agree for the ion distribution of an aqueous monovalent electrolyte 
near planar charged interfaces, as long as non-specific forces as well 
as excluded volume effects are negligible.
On the other hand, if the solution contains highly multivalent and/or large sized ions, 
then the Gouy-Chapman theory may severely qualitatively fail. 
A crucial missing ingredient in this theory is the inclusion of 
the finite size of the macroions that  can  lead to non-trivial phenomena, 
such as substrate's surface charge reversal, already with monovalent ions. 
\cite{Spitzer_JColIntfSci_1983,Kjellander_JCP_1998,Messina_EPL_2002,LozadaCassou_JCIS_2006} 
The lateral macroion-macroion electrostatic correlations that are also absent in a mean field theory
can be  attenuated (and even become marginal) at sufficiently high salt content, 
in contrast to excluded volume effects.  
In the past, Gonz{\'a}les-Mozuelos and Medina-Noyola \cite{GonzalesMozuelos_JCP_1991}
used integral equations to address the problem of macroions 
near repulsive/attractive charged walls interacting via an effective Yukawa potential.   
More recently, Netz  \cite{Netz_PRE_1999} investigated thoroughly and analytically 
the behavior of macroions near charged interfaces, based on the Debye-H\"uckel (DH) theory, 
but ignoring its finite size.
Thereby, the striking effect of surface charge amplification 
advocated here could not be captured by those approaches.
\cite{GonzalesMozuelos_JCP_1991,Netz_PRE_1999}
It is only very recently, that this phenomenon was reported by
Lozada-Cassou and coworkers. \cite{LozadaCassou_JCIS_2006,LozadaCassou_JPCB_2004}
For a size-asymmetrical electrolyte, 
\cite{LozadaCassou_JCIS_2006}
surface charge amplification was identified by solving numerically the modified Gouy-Chapman 
(where cations and anions have different distances of closest approach to the interface).
Intimately related to our work, the phenomenon of surface charge amplification
in presence of macroions was also found by applying a sophisticated hypernetted 
chain/mean spherical approximation (HNC/MSA) integral equation.
\cite{LozadaCassou_JPCB_2004}

\begin{figure}[b]
\includegraphics[width = 8.0 cm]{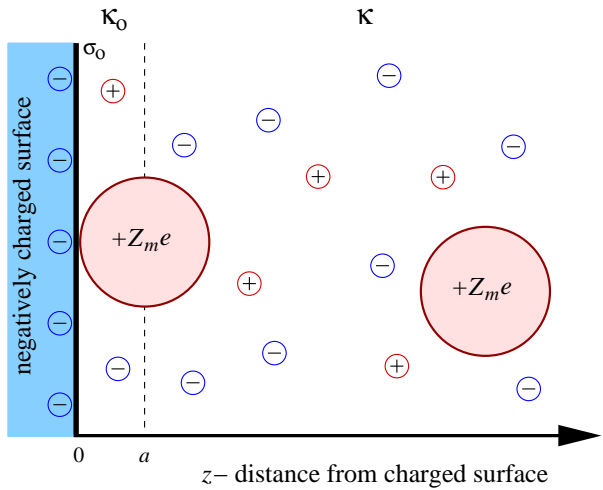}
\caption{
Schematic view of our model setup. The macroions are characterized
by a distance of closest approach $z=a$ to the charged surface, 
leading to two screening strengths 
$\kappa_0$ and  $\kappa$ for $0<z<a$ and $z>a$, respectively.
}
\label{fig.model_setup}
\end{figure}

In this work, we present a simple model, where the finite size of the macroions   
is taken into account, to reveal the mechanisms governing macroion adsorption. 
Although our approach is less accurate than the one used by Lozada-Cassou and coworkers,
\cite{LozadaCassou_JPCB_2004}
it presents the nice advantage to be analytical and very intuitive.
The basic driving force of surface charge amplification is that (spherical) macroions 
tend to be surrounded by its counterions over {\it its whole surface in a uniform manner}. 
This corresponds actually to the old classical 
Thomson's sphere problem. \cite{Thomson_PhilMag_1904}
As long as the strength of the surface charge density of the oppositely charged
substrate is low enough, a finite number of counterions of the macroions should    
stay in the vicinity of the interface (see Fig. \ref{fig.model_setup}),
leading to a surface charge amplification.
Our paper is organized as follows: Our (modified) DH theory 
is explained in Sec.  \ref{ sec.DH-theory}. The results are presented
in Sec. \ref{sec.results} and followed by a brief summary 
(see Sec. \ref{sec.conclu}).

\section{Debye H\"uckel theory \label{ sec.DH-theory}}

\subsection{Linearized Poisson-Boltzmann equation}

Our electrostatic model (see Fig. \ref{fig.model_setup}) 
resembles that employed by Spitzer \cite{Spitzer_JColIntfSci_1983}
who studied monovalent size-asymmetrical ions near a wall.
The negative substrate's surface charge density is denoted 
by $\sigma_0$.  
The macroions carry a positive central charge $+Z_me$ with $e$ representing 
the usual elementary charge and $Z_m$ its valency.
Excluded volume effects are taken into account via the distance of closest approach
$a$ to the charged interface (see Fig. \ref{fig.model_setup}).
Our model corresponds somehow to the minimal correction for macroion' size effect.  
Having to deal with a {\it finite} concentration of macroions at contact with a reservoir,
whose bulk value is given by $n_m$, electroneutrality requires a bulk macroion's 
counterion concentration $Z_m n_m$ (assuming monovalent point-like anions, i.e. $Z_-=-1$).
Those anions will be referred to as the {\it coions} of the substrate.  
Additional (point-like) monovalent counterions ($Z_+=+1$) 
and coions are also considered with a bulk salt concentration $n_s$.
%

An intuitive and widely used way to connect self-consistently 
the electrostatic potential, $\Psi (z)$, to the total charge density, 
$\rho(z) \approx \sum_{\alpha} n_{\alpha} Z_{\alpha} e \exp(-\beta Z_{\alpha} e \Psi)$
(with $\alpha$ standing for the ionic species), 
is provided by the so-called Poisson-Boltzmann equation that reads:
%
\begin{eqnarray}
\label{eq:PB}
\Delta \Psi (z) & =  &
-\frac{e}{\varepsilon_0 \varepsilon_r} 
\left[
-2n_s \sinh(e\beta \Psi) - Z_m n_m \exp(e\beta \Psi) 
\right. 
\nonumber \\ && \left.
+ Z_m n_m \exp(-Z_me\beta \Psi) \Theta(z-a)
\right],
\end{eqnarray}
%
where $\varepsilon_0$ ($\varepsilon_r$) is the vacuum (relative) permittivity, 
$\beta=1/(k_BT)$ is the reduced inverse temperature, and $\Theta$ is
the usual step (or Heaviside) function.
This non-linear differential equation \eqref{eq:PB} can only be solved numerically.
However a linearization of Eq. \eqref{eq:PB} 
[i.e., DH approximation],
valid for $|e\beta \Psi| \ll 1$ when $0<z<a$ and for
$|Z_me \beta \Psi| \ll 1$  when $z>a$, permits an analytical treatment
that is going to be discussed.

For the first diffuse region ($0<z<a$) made up uniquely of the little monovalent ions
(see also Fig. \ref{fig.model_setup}), the DH equation is given by
$\Delta u(z)  =  (\kappa_s^2  + \kappa_c^2)u + \kappa_c^2$
with
$\kappa_s^2 \equiv 8\pi \ell_B  n_s$ and
$\kappa_c^2 \equiv 4\pi \ell_B Z_m n_m$,
where we have introduced the Bjerrum length 
$\ell_B=\frac{e^2}{4\pi \varepsilon_0 \varepsilon_r k_BT}$
and the dimensionless variable  $u=e \beta \Psi$.
The corresponding solution reads:
%
\begin{eqnarray}
\label{eq:u_<}
u(z) = C_1 e^{-\kappa_0 z} +  C_2 e^{\kappa_0 z} - \frac{\kappa_c^2}{\kappa_0^2}
\quad (0<z<a),
\end{eqnarray}
%
where $\kappa_0^2 \equiv \kappa_s^2 + \kappa_c^2 = 4\pi \ell_B (2n_s+Z_m n_m)$.
$C_1$ and $C_2$ are integration constants that are going to be determined
after having applied the suitable boundary conditions.

For the second diffuse region ($z>a$) containing all the ions (including the macroions
- see also Fig. \ref{fig.model_setup}) the DH equation reads
$\Delta u(z) = \kappa^2 u$
with $\kappa^2 \equiv \kappa_0^2 + Z_m\kappa_c^2 = 4\pi \ell_B [2n_s+ Z_m(Z_m+1)n_m]$.
The physically sound solution with vanishing electric field
at $z \to +\infty$ corresponds to:
%
\begin{eqnarray}
\label{eq:u_>}
u(z) = C_3 e^{-\kappa z}
\quad (z>a),
\end{eqnarray}
%
where $C_3$ is a third integration constant.

%
\subsection{Boundary conditions}

Our electrostatic model system is completely characterized once the integration constants
$C_1$, $C_2$ and $C_3$ appearing in Eqs. \eqref{eq:u_<} and \eqref{eq:u_>} are specified.
To do so, we apply the three following boundary and/or matching conditions:
(i) 
The Gauss' law applied at the charged interface $z=0$ requires 
$u'(0) = \frac{2}{b}$, 
where $b=\frac{e}{2\pi\ell_B|\sigma_0|}$ is the so-called Gouy-Chapman length.
(ii) 
The zero intrinsic surface charge at $z=a$ 
imposes the continuity of the electric displacement 
$\vec D = \varepsilon_0 \varepsilon_r \vec E$ and hence also that of 
the electric field $\vec E$ [i.e., $u'(z \to a^-) = u'(z \to a^+)$]. 
\cite{boundary_electroneutrality}
(iii)
The continuity of the electrostatic potential at $z=a$ 
requires $u(z \to a^-)  =  u(z \to a^+)$.
%
%
The resulting set of three equations can be readily solved and yields:
\begin{eqnarray}
\label{eq:C_1}
\displaystyle 
C_1  & = & 
\displaystyle 
\frac{ -2 Q \displaystyle 
- \frac{2}{\kappa_0b} e^{\kappa_0 a} + \frac{\kappa_c^2}{\kappa_0^2}}
     {2 \cosh(\kappa_0 a)},
\\
\nonumber \\  
\label{eq:C_2}
\displaystyle 
C_2 & = &
\displaystyle 
\frac{ -2Q 
- \frac{2}{\kappa_0b} e^{\kappa_0 a} + \frac{\kappa_c^2}{\kappa_0^2}}
     {2 \cosh(\kappa_0 a)}
+ \frac{2}{\kappa_0b},
\\ 
\nonumber \\  
\label{eq:C_3}
C_3  & = & 
\displaystyle 
  -2e^{\kappa a}
\underbrace{
\left[ \frac{
  \frac{1}{\kappa_0b \cosh{(\kappa_0 a)}} + \frac{\kappa_c^2}{2\kappa_0^2} \tanh{(\kappa_0 a)}}
  { \frac{\kappa}{\kappa_0} + \tanh{(\kappa_0 a)}} 
  \right]}_{\equiv Q}.
\end{eqnarray}
%
%
Note that in the limit of point-like macroions ($\kappa_0 a \to 0$) one
recovers the well-known result $C_3 \to -\frac{2}{\kappa b}$.
At this stage all the relevant observables of the system can be in principle obtained within
the framework of the DH theory. 

\section{Results and discussion \label{sec.results}}

A pertinent quantity that is appropriate to characterize the strength 
of the macroion adsorption is provided by the 
contact potential of interaction $U_m$.
The latter corresponds to the external work accomplished upon bringing 
a macroion from infinity  ($z=+\infty$) to contact ($z=a$):
$U_m=-\int_{+\infty}^{a} Z_me E_z dz$ 
(with $E_z=-\frac{\partial \Psi}{\partial z}$ 
denoting the $z$-component of the electric field).
With the help of Eq. \eqref{eq:u_>} we get
$\beta U_m=Z_mu(a)=Z_mC_3e^{-\kappa a}$.
%
Using then Eq. \eqref{eq:C_3} for $C_3$, 
we obtain the following expression for $U_m$:
%
\begin{eqnarray}
\label{eq:Um_tanh}
\beta U_m & = & 
- \displaystyle 
   2Z_m
\left[ \frac{
  \frac{1}{\kappa_0b \cosh{(\kappa_0 a)}} + \frac{\kappa_c^2}{2\kappa_0^2} \tanh{(\kappa_0 a)}}
  { \frac{\kappa}{\kappa_0} + \tanh{(\kappa_0 a)}} 
  \right].
\end{eqnarray}
%
As can be seen from Eq. \eqref{eq:Um_tanh},
$U_m$ depends on many parameters, such as $Z_m$, $b$, $a$,
$\kappa$ and $\kappa_0$, \cite{kappa_c} making its understanding
rather difficult. Nonetheless, Eqs. \eqref{eq:u_<} - \eqref{eq:Um_tanh} 
suggest that more insight on the adsorption behavior can be gained by considering two limits:
(i) the high screening regime ($\kappa_0a \gg 1$) and (ii) 
the weak screening regime ($\kappa_0a \ll 1$).  
Without loss of generality, we are going to explore these two limits.

\subsection {High screening regime}

In the {\it high screening regime} ($\kappa_0 a \gg 1$), 
Eq. \eqref{eq:Um_tanh} can be approximated by:
%
\begin{eqnarray}
\label{eq:Um_high_screening}
\beta U_m & \simeq & 
1 - \frac{\kappa}{\kappa_0},
\end{eqnarray}
%
where the relation 
$\kappa_c^2=\frac{1}{Z_m}(\kappa^2-\kappa_0^2)$ has been used.
Interestingly, in this regime, $U_m$ depends only on the 
{\it screening strength contrast} $\kappa/\kappa_0$. 
One can also conveniently express $U_m$ as a
function of the bulk concentrations $n_m$ and $n_s$, which reads
%
\begin{eqnarray}
\label{eq:Um_high_screening_nm_ns}
\beta U_m & \simeq & 
1 - \sqrt{ \frac{2n_s+Z_m(Z_m+1)n_m}{2n_s+Z_mn_m}}.
\end{eqnarray}
%
\begin{figure}
\includegraphics[width = 8.0 cm]{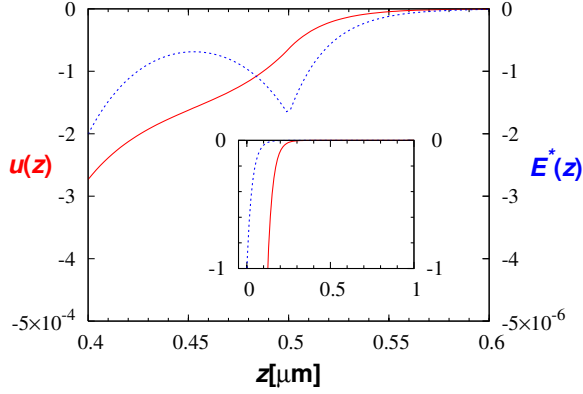}
\caption{
Profiles of the reduced electrostatic potential $u(z)$ (solid lines) and the reduced electric field 
$E^*(z)=-\frac{b}{2} u'(z)$ (dotted lines) around contact ($z=a=0.5 {\rm \mu m}$) calculated from 
Eqs. \eqref{eq:u_<} and \eqref{eq:u_>}.
The inset corresponds to a wider $z$-range.
The following parameters were chosen: 
$b=10^{-3} {\rm \mu m}$, $\ell_B=7.14 \times 10^{-4} {\rm \mu m}$, 
$Z_m=10^4$, $n_s=10^{-4} {\rm M}$, and $\phi_m = 10^{-3}$.
}
\label{fig.u_E_high_scr}
\end{figure}
%
Equation \eqref{eq:Um_high_screening_nm_ns} shows that 
$U_m$ is even independent of the Bjerrum length $\ell_B$,
which means that the effective wall-macroion attraction is 
entropically (or {\it depletion}) driven. 
To better understand this phenomenon, we have sketched on
Fig. \ref{fig.u_E_high_scr} the profile of $u(z)$ as well as profile of 
the reduced (dimensionless) electric field $E^*(z) \equiv -\frac{b}{2}u'(z)$ 
[with $E^*(z=0)=-1$] for a set of typical parameters
of charged colloidal suspensions: 
$a=0.5 {\rm \mu m}$, $b=10^{-3} {\rm \mu m}$ 
(corresponding to $|\sigma_0| \approx 0.036 {\rm Cm^{-2}}$ for an aqueous solvent), 
$Z_m=10^4$, $n_s=10^{-4}{\rm M}$,
a (fictive) equivalent volume fraction
 $\phi_m \equiv n_m \frac{4}{3} \pi a^3$ set to $10^{-3}$, 
(i.e.,  $n_m \approx 3.1714 \times 10^{-12} {\rm M}$  with M standing for mole per liter units).
Thereby we have $\kappa_0 a \simeq 16.4$.
The profile of $E^*(z)$ reveals an {\it unusual non-monotonic} behavior near contact.
This is a direct consequence of an accumulation of ``{\it excess}'' 
coions in the macroion depleted zone ($z<a$) 
leading to a {\it weaker} screening. 

\begin{figure}
\includegraphics[width = 8.0 cm]{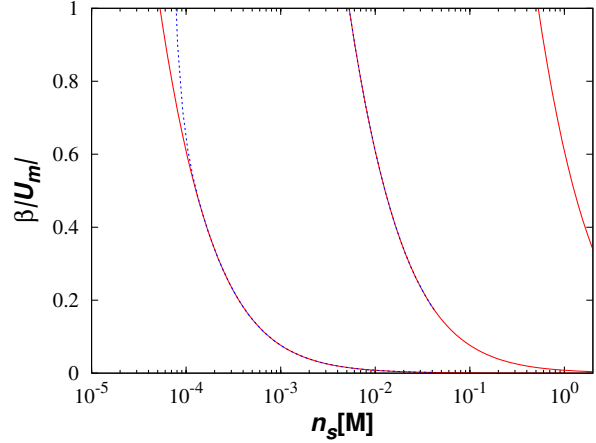}
\caption{
Reduced contact potential of interaction $\beta |U_m|$ as a function of the
molar bulk salt concentration $n_s$ in the strong screening regime
with $a=0.5 {\rm \mu m}$ and  $\phi_m = 10^{-3}$.
From left to right: $Z_m=10^4,10^5$ and $10^6$.
The solid lines were generated using Eq. \eqref{eq:Um_high_screening_nm_ns}.
The dotted lines were generated using Eq. \eqref{eq:Um_tanh} with 
$b=1 {\rm nm}$ and $\ell_B=0.714 {\rm nm}$.
For $Z_m=10^6$ the two curves are in practice identical.
}
\label{fig.Um_ns_high_scr}
\end{figure}

In order to further characterize the adsorption behavior in the high screening regime,   
we examine $U_m$ as predicted by 
Eq. \eqref{eq:Um_tanh} and Eq. \eqref{eq:Um_high_screening_nm_ns}
for some typical experimental values of the parameters $\kappa$
and $\kappa_0$ of charged colloidal systems.  
In Fig. \ref{fig.Um_ns_high_scr}, $-\beta U_m=\beta |U_m|$ 
is plotted against the bulk salt concentration $n_s$ 
for different prescribed values of $Z_m=10^4,10^5$ and $10^6$. 
Only values of  $\beta |U_m|$ smaller than unity are shown such as to
explore its behavior where the DH approximation is valid.  
Figure \ref{fig.Um_ns_high_scr} shows that beyond a certain threshold
of  salt content $n_s$ that is $Z_m$-dependent, 
no adsorption occurs as signaled by a nearly zero value of $U_m$.
At relatively low enough salt concentration, adsorption is favored where
$U_m$ increases with growing $Z_m$. 
The latter point can be clearly understood by noticing that for typical values 
of the parameters of charged colloidal systems ($Z_m\frac{n_m}{2n_s} \ll 1$ and 
$Z_m \gg 1$) we have 
$\beta U_m \approx1 - \sqrt{1+Z_m^2 \frac{n_m}{2n_s}}$.
Moreover, in the limit $Z_m^2 \frac{n_m}{2n_s} \ll 1$, the simple following relation 
$\beta U_m \approx -Z_m^2 \frac{n_m}{4n_s}$  holds. 


\subsection {Weak screening regime}

We now address the {\it weak screening regime} characterized by $\kappa_0 a \ll 1$.
In this situation Eq. \eqref{eq:Um_tanh} becomes:
%
\begin{eqnarray}
\label{eq:Um_weak_screening_final}
\beta U_m & \simeq & 
- \frac{2Z_m}{\kappa b} 
\left[ 1  
+ \frac{\kappa_0}{\kappa} \left\{ \frac{\kappa b}{2Z_m} 
  \left( \frac{\kappa^2}{\kappa_0^2} - 1\right) - 1 \right\} 
\kappa_0 a \right],
\nonumber \\
\end{eqnarray}
where the dispersion relation $\kappa^2 = \kappa_0^2 + Z_m \kappa_c^2$
has been used.
%
%
%
Equation \eqref{eq:Um_weak_screening_final} reveals that $U_m$ varies
affinely with macroion size $a$.
Depending on the sign of 
$ \frac{\kappa b}{2Z_m}\left( \frac{\kappa^2}{\kappa_0^2} - 1\right) - 1$,
$U_m$ may either decrease or increase with $a$.
For $\frac{\kappa b}{2Z_m} \gg 1$ we obtain the following limit behavior:
%
\begin{eqnarray}
\label{eq:Um_weak_scr_inf_b}
\lim_{\frac{\kappa b}{2Z_m} \to \infty} \beta U_m & \approx & 
- \left( 1 - \frac{\kappa_0^2}{\kappa^2} \right) \kappa a. 
\end{eqnarray}
%
In this ``entropic'' limit $|U_m|$ increases linearly with $a$,
meaning that a finite wall-macroion attraction persists even at vanishing 
surface charge density $\sigma_0$ due to excluded volume effect ($a \neq 0$). 
Noticing that 
%
%
\begin{eqnarray}
\label{eq:E}
E^*(a)=\frac{\kappa b}{2Z_m} \beta U_m,
\end{eqnarray}
%
we deduce that the strength of the electric field at contact  
$|E^*(a)|$ becomes larger than unity 
when  $\frac{\kappa b}{2Z_m}$ is sufficiently large 
[especially true for Eq. \eqref{eq:Um_weak_scr_inf_b}]. 
This feature corresponds to an 
{\it electric field (or surface charge) amplification} 
at contact.

%
\begin{figure}
\includegraphics[width = 8.0 cm]{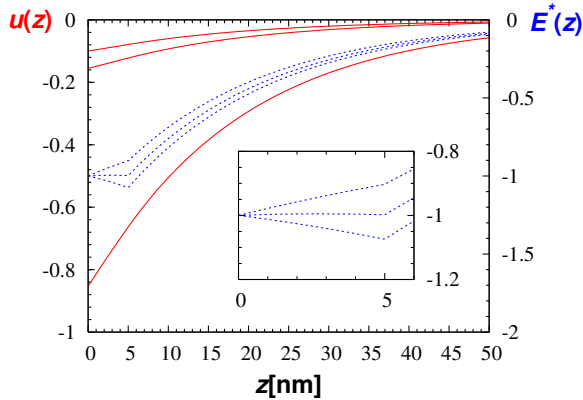}
\caption{
Profiles of the reduced electrostatic potential $u(z)$ 
(solid lines - from top to bottom: $b=500, 300, 50  {\rm nm}$) 
and the reduced electric field 
$E^*(z)=-\frac{b}{2} u'(z)$ 
(dotted lines - from top to bottom: $b=50, 300, 500  {\rm nm}$). 
The inset corresponds to a magnification of $E^*(z)$. 
The following parameters were chosen: 
$a=5 {\rm nm}$, $\ell_B=0.714 {\rm nm}$, 
$Z_m=10$, $n_s=10^{-4} {\rm M}$, and $\phi_m = 10^{-3}$.
}
\label{fig.u_E_weak_scr}
\end{figure}
%

Let us consider some typical experimental values of parameters 
representative of charged {\it micellar} systems
(see caption of Fig. \ref{fig.u_E_weak_scr}).
Thereby we have $\kappa_0a \simeq 0.177$.
To access the mechanisms of wall-macroion attraction in the weak screening regime,
we have plotted the profiles of  $u(z)$ and $E^*(z)$ in Fig. \ref{fig.u_E_weak_scr}
for various values of $b$.
Strikingly, in the macroion depleted zone, the electric field gets monotonically 
{\it amplified}  (from $z=0$ up to contact $z=a=5 {\rm nm}$) at (very) large $b$ 
[corresponding to poorly charged interfaces 
($|\sigma_0| \lesssim 10^{-4} {\rm Cm^{-2}}$) - see Fig. \ref{fig.u_E_weak_scr}]. 
This phenomenon is again due to an accumulation of ``excess'' 
coions in the macroion depleted zone,
that leads here to a  net surface charge that is more negative than $\sigma_0$.
It is important to remark, that surface charge amplification does not
necessarily involve a  subsequent charge reversal, as clearly
indicated in Fig. \ref{fig.u_E_weak_scr}.
In fact this charge-amplification phenomenon was already observed, although uncommented,
in computer simulations 
[see Fig. 4(b) in Ref. \cite{Tanaka_JCP_2001}
and Fig. 3 in Ref. \cite{Maiti_NanoLett_2006}].
%
%

\section{Concluding remarks \label{sec.conclu}}

To summarize, we have studied analytically within the framework of the Debye-H\"uckel theory
the (weak) electrostatic adsorption of macroions
at oppositely charged planar surfaces. Taking into account the crucial role of the 
finite size of the macroions,
our model reveals non trivial adsorption driving forces.
In the strong screening regime, the wall-macroion attraction strength at contact is exclusively
governed by the screening strength contrast $\kappa/\kappa_0$. 
In the weak screening regime, the wall-macroion attraction strength can either decrease or 
increase with macroion size, depending on the surface charge density of the substrate.
In particular, at sufficiently small surface charge densities an effective electric filed 
(or equivalently an effective surface charge) amplification sets in.
All these adsorption mechanisms have a common feature, namely, the accumulation of excess coions in the
macroion depleted zone.
The latter mechanism also explains the surface charge amplification recently reported by 
molecular dynamics simulations for DNA-dendrimer complexation. \cite{Maiti_NanoLett_2006}
Our findings could be experimentally verified by employing an (extra) ultra fine dispersion  
of charged particle tracers (with a smaller size than the macroions and such as to nearly not modify
the screening strengths $\kappa$ and $\kappa_0$), whose density 
profile should reveal the electric filed' one in the macroion depleted zone. 

\acknowledgments 
The author thanks M. Lozada-Cassou for enlightening discussions.
Financial support from DFG via LO418/12 and SFB TR6 is acknowledged.


\begin{thebibliography}{10}
\bibitem{boundary_electroneutrality}
Note that the same result is obtained by requiring global elecroneutrality:
  $\int_0^{\infty} \sum_{\alpha} \rho_{\alpha}(z) dz = - \sigma_0$.
\end{thebibliography}

\end{document}